\documentclass[twocolumn,showpacs,english,superscriptaddress,preprintnumbers,amsmath,amssymb,floatfix]{revtex4-1}

\usepackage[T1]{fontenc}
\usepackage[latin9]{inputenc}
\usepackage{dcolumn}
\usepackage{bm}
\usepackage{graphicx}
\usepackage{color}
\usepackage{esint}
\usepackage{babel}
\usepackage{amsfonts}
\usepackage{slashed}
\usepackage{enumerate}

\begin{document}

\title{Majorana qubits in topological insulator nanoribbon architecture}

\author{J. Manousakis}
\affiliation{Institut f\"ur Theoretische Physik,
Universit\"at zu K\"oln, Z\"ulpicher Str.~77, D-50937  K\"oln, Germany}
\author{A. Altland}
\affiliation{Institut f\"ur Theoretische Physik,
Universit\"at zu K\"oln, Z\"ulpicher Str.~77, D-50937  K\"oln, Germany}
\author{D. Bagrets}
\affiliation{Institut f\"ur Theoretische Physik,
Universit\"at zu K\"oln, Z\"ulpicher Str.~77, D-50937  K\"oln, Germany}
\author{R. Egger}
\affiliation{Institut f\"ur Theoretische Physik,
Heinrich-Heine-Universit\"at, D-40225  D\"usseldorf, Germany}
\author{Yoichi Ando}
\affiliation{Physics Institute II,
Universit\"at zu K\"oln, Z\"ulpicher Str.~77, D-50937  K\"oln, Germany}
\date{\today}

\begin{abstract}
We describe designs for the realization of topological Majorana qubits in terms of
proximitized topological insulator nanoribbons pierced by a uniform axial magnetic
field.  This platform holds promise for particularly robust Majorana bound states,
with easily manipulable inter-state couplings. 
We propose proof-of-principle experiments for initializing, manipulating, 
and reading out Majorana box qubits defined in floating
devices dominated by charging effects. We argue that the platform offers
design advantages which make it particularly suitable for extension to qubit network
structures realizing a Majorana surface code.
\end{abstract}

\maketitle

\section{Introduction}\label{sec1}

Majorana  fermions are currently becoming a reality as 
emergent quasiparticles in topological superconductors, see
Refs.~\cite{Hasan2010,Zhang2011,Alicea2012,Leijnse2012,Tanaka2012,
Beenakker2013,Ando2015,Sato2016} for reviews. In topological superconductors, the
pairing of effectively spinless fermions implies that  quasiparticles at positive and
negative energies are related by Hermitean conjugation and, as a consequence,
Majorana fermions emerge at zero energy. If the ensuing states are localized in space,
they define  Majorana bound states (MBSs). The operators corresponding to MBSs 
are self-adjoint, $\gamma=\gamma^\dagger$, i.e., particle
and antiparticle are identical, and anticommute with all other fermion operators. Two
Majoranas $\gamma_1$ and $\gamma_2$ may be combined to an ordinary fermion, $c=(\gamma_1+i\gamma_2)/2$.  For spatially well separated MBSs,
$c$ describes a zero-energy fermion state and the ground state of the topological superconductor will be degenerate with respect to even/odd fermion parity. 
The extension to $2N$ separated MBSs gives rise to a 
$2^N$-fold degenerate ground-state manifold. These ground states are candidates for applications in quantum information processing (QIP), 
where the spatial separation between MBSs provides a topological
protection mechanism against decoherence \cite{Nayak2008}. 

Most proposals for implementing Majorana-based QIP rely on non-Abelian braiding
operations in the degenerate ground-state manifold~
\cite{Nayak2008,Fu2008,Sau2010,Alicea2011,Clarke2011,Hyart2013,Vijay2016,Aasen2016}.
A recent alternative approach suggests the engineering of patterns of low-capacitance
mesoscopic superconducting islands harboring
MBSs~\cite{Beri2012,Altland2013,Beri2013,Plugge2017,Karzig2017}, where the ensuing
two-dimensional (2D) Majorana surface code
\cite{Terhal2012,Vijay2015,Landau2016,Plugge2016} defines a topologically ordered
Abelian state of matter. The fundamental design advantage of the surface code is that
only modest fidelities $\approx 0.99$ are required for elementary gate operations \cite{Terhal2015}.
However, all approaches to Majorana-based QIP have in common that they rely  on the
realizability of robust and easily manipulable MBSs, which in turn define the
hardware qubits of the corresponding architecture.   In particular, one must be able
to initialize, manipulate, and read out the corresponding qubit states in a
phase-coherent environment while facing the challenge of scalability to 2D extended
structures.

Currently, two platforms are intensely studied and hold promise to meet these
criteria. The first  builds on spin-orbit-coupled semiconductor (InAs or InSb)
nanowires proximitized by $s$-wave superconductors (Al or NbTiN), where evidence for
MBS formation has already been seen in zero-bias conductance peaks
\cite{Mourik2012,Leo2017} and in Coulomb blockade spectroscopy
\cite{Albrecht2016,Albrecht2016b}. The second platform employs 
1D edge states of the layered quantum spin Hall insulator HgTe proximitized by Nb
\cite{Deacon2017,Bocquillon2017}. Both platforms offer specific advantages but also face specific
challenges.  For instance, while the semiconductor approach benefits from decades of experience in device technology it is confined to MBSs realized in  narrowly
defined parameter regimes close to the bottom of a semiconductor band
\cite{Alicea2012}. 

In this paper we propose an alternative Majorana qubit architecture and 
outline how to implement basic QIP operations in it.
 Our setup employs nanoribbons of 3D topological insulator (TI) materials, e.g.,  
Bi$_2$Se$_3$ or Bi$_2$Te$_3$, proximitized by conventional $s$-wave superconductors.
Surface states of proximitized TIs are expected to realize 
topological superconductors \cite{Fu2008}, and theoretical work has 
predicted the formation of MBSs  near the ends of such ribbons \cite{Cook2011,Cook2012,Sitthison2014,Juan2014,Ilan2014,Huang2016}. 
Since these MBSs are built from protected surface states of a bulk topological insulator,
they are expected to show high levels of robustness. 
Although  no experimental evidence for MBSs in this material class has been reported yet, we are positive that there is no fundamental obstacle preventing success 
along that direction. Below we will describe a tunable Majorana qubit realization 
using this platform and outline how to perform simple quantum operations with it.
 
The  layered structure of most TIs implies that 1D nanowires formed from such
materials grow in a tape-like shape. For the ensuing nanoribbons, rather small cross
sections ($\approx 40\times 100$~nm$^2$ \cite{Cho2015,Jauregui2016}) are feasible,
where surface states located on opposite sides of the nanoribbon still have a finite
overlap.  The corresponding surface state bands are inside the bulk TI energy gap
of width $0.3$~eV \cite{Hasan2010}. In general, however, they
exhibit a finite-size gap because of this overlap.
Remarkably, the presence of an axial magnetic flux, $\Phi$, equal  to half a flux
quantum, $\Phi=\Phi_0/2\equiv h/2e$, may close this finite-size gap. In practice, the
application of a magnetic field of order $0.5$~T is expected to generate a single
gapless 1D mode
\cite{Gornyi2010,Bardarson2010,Zhang2010,Egger2010,Kundu2011,Bardarson2013}. This
helical 1D mode is insensitive to elastic impurity scattering and 
experimental efforts towards the confirmation of its existence have been made 
\cite{Cho2015,Jauregui2016}.  Once a TI nanoribbon with
$\Phi=\Phi_0/2$ is proximitized by an $s$-wave superconductor, a 1D topological
superconductor phase with Majorana end states emerges
\cite{Cook2011,Cook2012,Sitthison2014,Juan2014,Ilan2014,Huang2016}. Conceptually,
these MBSs form under conditions of symmetry class $D$, where spin SU$(2)$ symmetry,
time reversal symmetry, and particle number conservation are all broken
\cite{footnote1}, and they are robust against both conventional and pair-breaking
disorder. Importantly, they also tolerate arbitrary chemical potentials located in
the bulk band gap.  The above features indicate that the TI nanoribbon platform lends itself to robust Majorana qubit implementations and,  eventually, for
QIP applications.

\begin{figure}[t]
\centering
\includegraphics[width=\columnwidth]{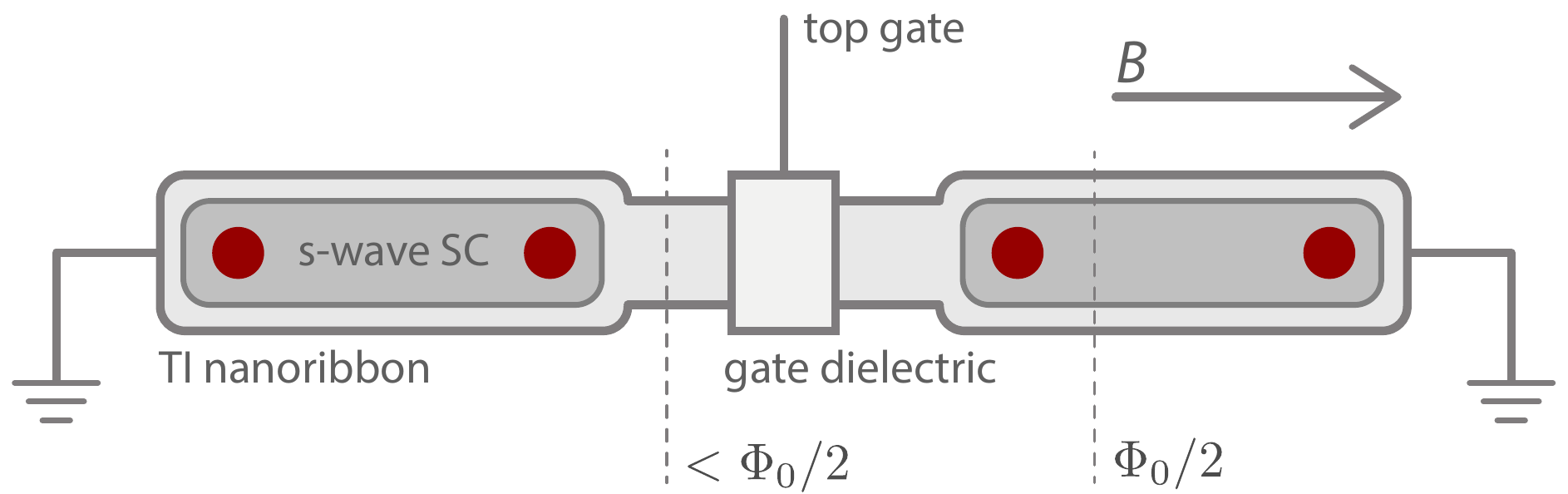}
\caption{ Schematic device with four Majorana states built from a 
TI nanoribbon in a uniform axial magnetic field $B$.  We assume that $B$ results 
in the flux $\Phi=\Phi_0/2$ through the (thick) outer parts which are proximitized 
by an $s$-wave superconductor (SC) layer. 
The 1D surface states in the non-proximitized narrow central part of the device are gapped since here the magnetic flux $\Phi$ is well below $\Phi_0/2$. 
The transparency of this junction can be tuned by an electrostatic top gate.  
Majorana end states are indicated as red dots. Note that this device is grounded. }
\label{figure1}
\end{figure}

Below we will consider various MBS and qubit designs building on the above
construction. We first consider a grounded setup comprising two proximitized wire
segments separated by a non-proximitized central region of narrower geometric cross
section, see Fig.~\ref{figure1} for a schematic illustration. The  narrowed central
region might be  fabricated by electron beam lithography and wet etching in order
to minimize defects. Its reduced cross section implies that it is threaded by a
magnetic flux lower than $\Phi_0/2$, and therefore exhibits a size quantization gap
in its surface state spectrum. Majorana bound states then form at the interfaces
between regions of different (superconducting vs size quantization) spectral gap
type, where the gate-tunable weak link represents a Josephson junction between two
topologically superconducting wires.
  
  As we will discuss in Sec.~\ref{sec3}, using either floating or grounded versions
 of the device in Fig.~\ref{figure1}, a variety of means to access and manipulate
 quantum states encoded by the four MBSs are available. The option to switch between
 floating and grounded versions by means of electrostatic gating makes it
 possible to employ the manipulation schemes proposed by Aasen \textit{et
 al.}~\cite{Aasen2016}. In particular, one may detect the occupation state of the
 fermion formed from the central MBS pair in Fig.~\ref{figure1} by the
 parity-to-charge conversion technique described in Ref.~\cite{Aasen2016}. In most of
 this paper, however, we pursue an alternative approach where floating
 devices and Coulomb blockade effects are crucial
 \cite{Albrecht2016,Beri2012,Altland2013,Beri2013,Fu2010, Hutzen2012}. In that case,
 electron transfer between different parts of the device is governed by non-local
 electron tunneling, and Majorana box qubits can be defined along the lines of
 Refs.~\cite{Plugge2017,Karzig2017}. These qubits may ultimately be arranged in a
 2D TI nanoribbon network in order to implement a Majorana surface code. (We note in
 passing that the alternative TI-based Majorana surface code proposal of
 Ref.~\cite{Vijay2015} operates in a rather different parameter regime, where the
 Josephson coupling between different qubits is essential.) However, although we will briefly sketch these long-term perspectives,  the main emphasis of this paper is
 on basic design aspects and suggesting proof-of-principle experiments testing the proposed topological qubits.

Before entering a detailed discussion, let us summarize
the structure of the remainder of this paper and offer 
guidance to the focused reader. 
In Sec.~\ref{sec2}, we provide a theoretical description of Majorana states
for the grounded TI nanoribbon device in Fig.~\ref{figure1}.  
To keep the presentation self-contained, Secs.~\ref{sec2a} and \ref{sec2b} also
summarize those results of Refs.~\cite{Cook2011,Cook2012,Sitthison2014,Juan2014,Ilan2014,Huang2016} 
 that are relevant to our subsequent discussion. 
In Sec.~\ref{sec2c}, we describe in detail how the hybridization energy 
corresponding to the overlap between the two central MBSs in Fig.~\ref{figure1} 
can be manipulated via suitable gate electrodes.  
A floating version of the device in Fig.~\ref{figure1}, where Coulomb charging effects
are important,  is then addressed in Sec.~\ref{sec3}.  In Sec.~\ref{sec3a}, we 
analyze effects introduced by the presence of a charging energy and/or Josephson couplings.  When the charging energy dominates, a device as sketched 
in Fig.~\ref{figure4}(b) can encode a Majorana box qubit.       
In Sec.~\ref{sec3b} and Sec.~\ref{sec3c}, we briefly review key ideas of 
Refs.~\cite{Plugge2017}  and \cite{Landau2016} concerning the device layout
and basic operation principles, and transfer those ideas to the
TI implementation.  A detailed comparison between our proposal and
alternative platforms is then provided in Sec.~\ref{sec3d}.
Finally, Sec.~\ref{sec4} concludes with an outlook, where we  sketch 
how a Majorana surface
code could be implemented by arranging such qubits in a network, cf.~Refs.~\cite{Landau2016,Plugge2016}.  Technical details 
concerning Sec.~\ref{sec2} have been delegated to Appendix~\ref{appa}. 

\section{Proximitized nanoribbon device} 
\label{sec2}

Let us now turn to a theoretical description of the basic device  
shown in Fig.~\ref{figure1}.  For this grounded device, it is sufficient
to study single-particle properties. However, 
Coulomb charging effects are crucial for floating devices and will be 
taken into account in Sec.~\ref{sec3}. 

\subsection{Model}\label{sec2a}

The TI nanoribbon containing a bottleneck in the central region
is modeled in terms of a long cylindrical TI nanowire along the $z$-direction
 with spatially varying radius $R(z)$. 
For most TI materials, nanoribbons naturally grow with a rectangular cross section.
However,  the low-energy band structure turns 
out to be very similar to the one found for cylindrical wires with the same cross section    \cite{Kundu2011,Cook2012,Bardarson2013}.  
The cylindrical geometry is technically easier to handle because of
azimuthal angular momentum conservation, where the 
 angular momentum quantum number $j$ is quantized in half-integer units. 

We model the central region by a simple step function profile of width $W$ centered around $z=0$,
\begin{equation}\label{radiusdep}
R(z) = \left\{ \begin{array}{cc} R_0,& |z|\le W/2, \\  R,& |z|> W/2 ,\end{array}\right.
\end{equation}
where $R_0<R$.  This step function modeling is  motivated by  convenience as it allows for simple analytical solutions via wave function matching.   
It assumes that interfaces extend over a few lattice spacings  (which in turn are $\approx 3$~nm
 \cite{Hasan2010}), since otherwise the low-energy approach used below is not applicable and TI states above the bulk gap can become important.
Smooth interfaces, where  $R(z)$ changes over longer scales,
can be described via slightly more involved solution schemes.
However, since eigenstates show the same asymptotic behavior  far away from 
the interfaces as for the step-like profile
\eqref{radiusdep}, we do not expect 
qualitatively different physics.

In the presence of a constant axial magnetic field $B$,  
the $z$-dependence of the nanowire radius implies a 
reduced magnetic flux through the central region.  
Defining the dimensionless magnetic flux 
$\varphi(z)=\Phi(z)/\Phi_0$, and assuming that the field strength $B$ 
has been adjusted to give $\varphi=1/2$ in the outer regions, we obtain
\begin{equation}\label{innerflux}
\varphi(z)= \frac{\Phi(z)}{\Phi_0}=\left\{
\begin{array}{cc} \varphi_0 \equiv  (R_0/R)^2/2, & |z|\leq W/2, 
\\ 1/2,& |z|>W/2,
\end{array}\right.
\end{equation}
with $\varphi_0<1/2$.
Equation \eqref{innerflux} neglects magnetic screening (flux channeling) in 
the central region. Such effects are expected to be tiny due to
the smallness of the magnetic susceptibility, in particular in the central region where 
size quantization implies a gap in the surface state spectrum (see below).

We next assume the presence of a complex-valued 
superconducting gap parameter $\Delta(z)$ introduced via the proximity to
$s$-wave superconductors in the outer regions of the device, see Fig.~\ref{figure1}. 
 For simplicity, we assume that the absolute value of the proximity-induced 
gap is identical on both sides, $|\Delta(z)|=\Delta$ for $|z|>W/2$. 
With the phase difference $\phi$ across the weak link, we have
\begin{equation}\label{deltadef}
\Delta(z)=\left\{ \begin{array}{cc}0, &|z|\leq W/2,\\
\Delta e^{i\phi/2}, & z<-W/2,\\ \Delta e^{-i\phi/2}, & z>W/2.
\end{array}\right.
\end{equation}
For a floating device, $\phi$ will be a dynamical quantity. 
We note that Eq.~\eqref{deltadef} does not take into account
rotational symmetry breaking by the $s$-wave superconductors. Such 
effects have been considered in Ref.~\cite{Sitthison2014}.

Finally, the top gate electrode in Fig.~\ref{figure1} induces an  
electrochemical potential $\mu(z)$ in the central region, 
where we have no $s$-wave superconductor and gating is possible.  
 Assuming a constant but tunable value for this potential, we obtain
\begin{equation}\label{delmu}
\mu(z)=\left\{ \begin{array}{cc}
\mu, & |z|\leq W/2,\\0, &|z|> W/2.
\end{array}\right.
\end{equation}
As detailed in Sec.~\ref{sec3d}, a finite value of $\mu$ 
in the region $|z|>W/2$ is not expected to cause qualitative changes.

Under the conditions defined above, the surface states of this TI nanowire may be computed via different methods, including microscopic tight-binding calculations or   
${\bf k}\cdot {\bf p}$ theory supplemented with Dirichlet boundary 
conditions on the surface, see, e.g., Ref.~\cite{Egger2010}.  
 However, as detailed in 
 Refs.~\cite{Cook2012,Gornyi2010,Zhang2010,Bardarson2010,Egger2010,
Kundu2011,Bardarson2013}, the results of such calculations
are well reproduced by a simple  description in terms of  
effectively 2D massless Dirac fermions wrapped onto the surface of the device, subject to the constraint that spin is oriented tangentially to the surface and
perpendicularly to the momentum at any point.   
The full surface state solution includes a radial part describing a
 rapid exponential decay into the bulk of the nanowire. 
However, this radial dependence of the wave functions will be 
left implicit throughout.

For fixed half-integer conserved angular momentum $j$,
the structure of a surface state in spin space is then given by
\begin{equation}\label{states}
\psi_j(z,\theta) = \frac{ e^{ij\theta} }{\sqrt{2\pi}} 
\left( \begin{array}{c} e^{-i\theta/2} f_j(z) \\ e^{i\theta/2} g_j(z) 
\end{array}\right),
\end{equation}
where the angle $\theta$ parametrizes the circumference of the nanowire and
the $z$-dependent functions $f_j$ and $g_j$ obey the normalization 
$\int dz (|f_j|^2+|g_j|^2)=1$.  We thus arrive at a reduced 1D formulation, where the Hamiltonian effectively acts on spinor states $( f_j(z) ,g_j(z) )^T$.
In the presence of a superconducting gap $\Delta(z)$,
surface states inside the bulk TI energy gap are then obtained as eigenstates of the 
Bogoliubov-de Gennes (BdG) Hamiltonian
\begin{eqnarray}\label{Ham0}
H_{\rm BdG} &=& \left( \begin{array}{cc} H_0(z) & i\sigma_y \Delta(z) \\
-i\sigma_y \Delta^*(z) & -H_0 (z) 
\end{array} \right) ,\\
\nonumber
H_0 &=& -i\hbar v_1 \sigma_y \partial_z - \frac{\hbar v_2}{R(z)}
\left[j+\varphi(z)\right] \sigma_z -\mu(z) \sigma_0 . 
\end{eqnarray}
For a detailed derivation, see Ref.~\cite{Cook2012}. 
Here Pauli matrices $\sigma_{x,y,z}$ (and identity $\sigma_0$) act in spin space,
and the explicit $2\times 2$ structure in Eq.~\eqref{Ham0} 
refers to particle-hole (Nambu) space. Moreover, $v_1$ and $v_2$ are Fermi velocities along the axial and circumferential direction, respectively, which depend on TI material parameters.  
Note that the magnetic flux effectively shifts the 
quantized angular momentum number $j\to j+\varphi(z)$.

In the absence of the constriction ($W= \phi=0$), and
assuming an infinitely long wire, the system is translationally
invariant and BdG eigenstates are plane waves with longitudinal momentum $k$.
The piercing of the system by a uniform flux $\varphi$ leads to the dispersion
relation \cite{Cook2012}
\begin{equation}\label{spectr0}
E_{k,j,\sigma,\sigma'} = \sigma\sqrt{(\hbar v_1 k)^2 + (M_j+\sigma' \Delta)^2 },
\end{equation}
with $\sigma,\sigma'=\pm$ and the size quantization gap parameters $M_j=
\hbar v_2 |j+\varphi| / R.$ In general, the spectrum in Eq.~\eqref{spectr0} is gapped
either by the size quantization gap ($M_j$) or by the
superconducting gap ($\Delta$).  However, a gapless branch exists for $M_j=\Delta$
and $\sigma'=-1$, and a topological phase transition is expected at this gap-closing
point \cite{Alicea2012}. On general grounds, this signals the formation of
Jackiw-Rossi zero modes corresponding to localized Majorana fermions at interfaces
between two regions dominated by different gap types.

We next note that for the  angular momentum mode $j=-1/2$, the
size quantization gap,
\begin{equation}\label{massgap}
    M_{-1/2}\equiv M(\varphi) =\frac{\hbar v_2}{R} |\varphi-1/2|,
\end{equation}
vanishes when the flux equals half a
flux quantum, $\varphi=1/2$. In that case, the wire supports a  gapless
helical 1D mode for $\Delta=0$,
cf.~Refs.~\cite{Gornyi2010,Zhang2010,Bardarson2010,Egger2010}. Below we 
will assume that the
cross section of the TI nanowire is small such that $\hbar v_2/R$ represents a large
energy scale. In that case, all modes with angular momentum $j\neq -1/2$ can be
neglected. We therefore retain only the $j=-1/2$ mode in what follows.
The effects of flux mismatch away from $\varphi=1/2$ 
 and physical mechanisms which may cause it are
addressed in Sec.~\ref{sec3d} below.

\subsection{Majorana states} \label{sec2b}
 
Turning back to the device in Fig.~\ref{figure1}, let us still assume an
infinitely long TI nanowire but now with a finite size $W$ of the central region
with narrowed cross section. The above discussion shows
that the outer parts are dominated by a superconducting gap $(M=0, \Delta\not=0)$.
On the other hand, in the central part, the superconducting gap
vanishes but we have a size quantization gap $(M_0\not=0,\Delta=0)$. 
Here we define $M_0\equiv M(\varphi_0)$ with $R\to R_0$
in Eq.~\eqref{massgap}, where $\varphi_0$ has been introduced in Eq.~\eqref{innerflux}.
The points $z=\pm W/2$ thus define interfaces where gaps of different nature 
meet each other. As a consequence, the solution of the BdG equation, $H_{\rm BdG}\Psi=E\Psi$,  must include MBSs localized near these two
 interface points. 
 
 In general, the leakage of Majorana wave functions into 
 the central region will then result in a finite hybridization energy $\varepsilon$,
where the two MBSs correspond to a zero-energy fermion state only 
for $\varepsilon\to 0$ while the degeneracy of both parity states is lifted otherwise. 
 We will quantitatively determine $\varepsilon$ and show that it 
can be efficiently tuned, e.g., by varying the electrochemical potential $\mu$.
Only moderate values $\mu<M_0$ are considered below since otherwise also
higher-energy modes with $j\ne -1/2$ have to be taken into account.

Away from the interfaces, $z\ne \pm W/2$, the problem is effectively uniform. 
The 1D BdG equation is then either solved by a plane wave ansatz (for high energies) or 
by an evanescent state ansatz (for small $|E|$). The  solutions in the sub-gap regime 
$|E|<\Delta$ are detailed in App.~\ref{appa}.
The requirement of continuity of the spinor wave function $\Psi(z)$ at the interface 
points implies that a corresponding determinant vanishes, 
\begin{equation}\label{deter}
D(E)=0,
\end{equation} 
where $D(E)$ is specified in Eq.~\eqref{Ddef}
 for $|E|<{\rm min}(\Delta,M_0)$ but otherwise arbitrary parameters. 
 The condition (\ref{deter}) determines the low-energy spectrum of the system,
 which for generic parameter sets is established by numerical solution.

\subsection{Hybridization between Majorana states}\label{sec2c}

A robust feature found by solving Eq.~\eqref{deter} is the existence of
sub-gap states at $E=\pm \varepsilon$, representing the expected pair of MBSs. Under the self-consistent assumption $|\varepsilon|\ll {\rm min}(\Delta,M_0)$, the value of $\varepsilon$ can be obtained from Eq.~\eqref{deter} 
by second-order expansion of $D(E)$  in $E$.  We find
\begin{equation}\label{4pi}
\varepsilon(\phi) = \varepsilon(0) \cos(\phi/2),
\end{equation}
where the result at phase difference $\phi=0$ is
\begin{equation}\label{exposcal}
\varepsilon(0) = \frac{2 \Delta}{M_0} 
\frac{(\hbar v_1/\xi)^2}{\Delta + \hbar v_1/\xi } e^{-W/\xi}
\end{equation}
with the length scale 
\begin{equation}\label{xidef}
\xi=\frac{\hbar v_1}{\sqrt{M_0^2-\mu^2}}.
\end{equation} 
Notice the $4\pi$-periodic behavior of $\varepsilon(\phi)$ in Eq.~\eqref{4pi}, which
is the periodicity shown by topological Josephson junctions \cite{Alicea2012} where
 contact between two superconductors is established by MBSs.
We mention in passing that  small $2\pi$-periodic admixtures add to Eq.~\eqref{4pi}
if higher-lying surface states with $j\ne -1/2$ are included.

\begin{figure}[t]
\centering
\includegraphics[width=0.95\columnwidth]{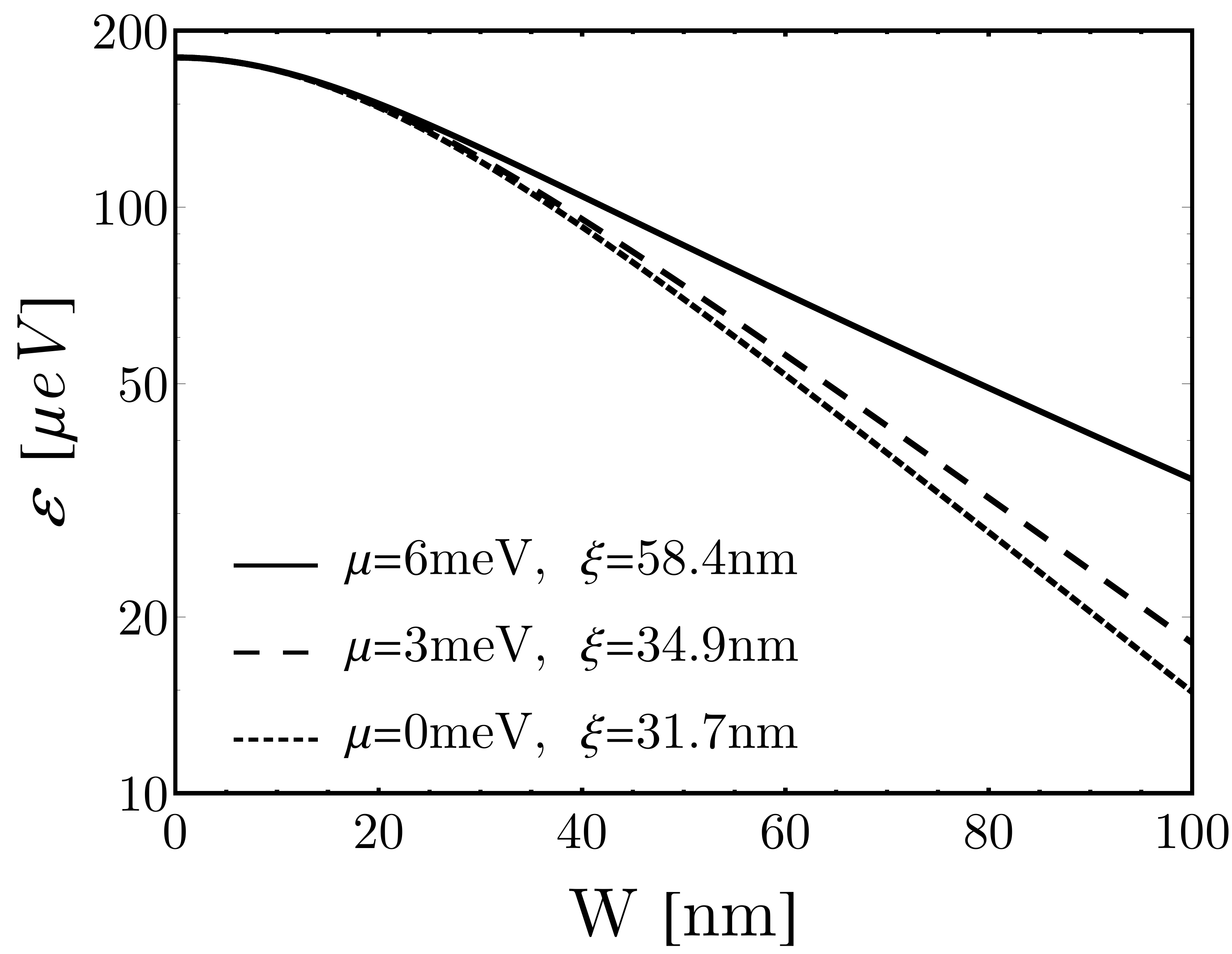}
\caption{Majorana hybridization energy $\varepsilon$ (in units of $\mu$eV) 
vs constriction length $W$ (in nm) for the device in Fig.~\ref{figure1}. 
Taking parameters for Bi$_2$Se$_3$ with
$\Delta=0.18$~meV, $M_{0}\simeq 7.14$~meV, and $\phi=0$, 
results are shown for several values of $\mu$. 
The quoted values for $\xi$ follow from Eq.~\eqref{xidef}.
For $W\gg \xi$, these semi-logarithmic plots are consistent with
$\varepsilon\sim e^{-W/\xi}$, see Eq.~\eqref{exposcal},
 while $\varepsilon\to \Delta$ for $W\to 0$.}
\label{figure2}
\end{figure}

For  $W\gg \xi$, the energy $\varepsilon$ in Eq.~\eqref{exposcal} becomes 
exponentially small, $\varepsilon\sim e^{-W/\xi}$, as
expected for the hybridization energy of far separated MBSs. 
 The explicit solution in App.~\ref{appa} shows that $\Psi(z)$ has an exponential 
 decay away from the interface points into the
 proximitized parts ($|z|>W/2$) on the length scale $\hbar v_1/\Delta$, 
 see Eq.~\eqref{eqA1}. Similarly, the length scale $\xi$ in Eq.~\eqref{xidef} 
governs the decay of $\Psi(z)$ into the central part,
see Eq.~\eqref{eqA2}.  For a constriction of length $W\gg \xi$, the Majorana overlap 
thus becomes exponentially small and we encounter a pair of Majorana zero modes.

\begin{figure}[t]
\centering
\includegraphics[width=0.9\columnwidth]{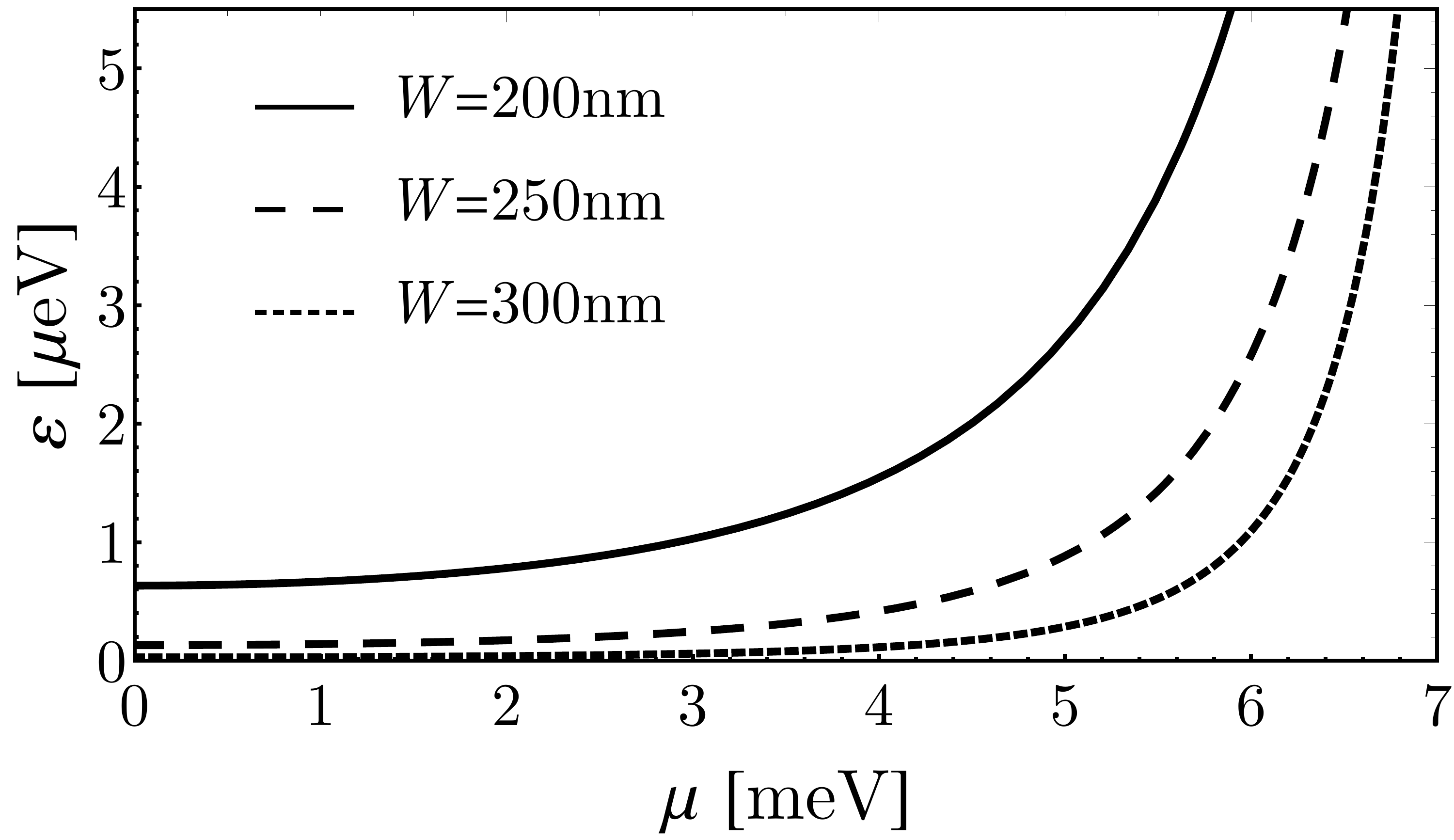}
\caption{Hybridization energy $\varepsilon$ (in $\mu$eV) vs 
electrochemical potential $\mu$ (in meV) for several values of the 
constriction length $W$. Parameters are as in Fig.~\ref{figure2}, i.e.,
for Bi$_2$Se$_3$ with $\Delta=0.18$~meV, $M_0\simeq 7.14$~meV, and $\phi=0$.
\label{figure3}}
\end{figure}

We next discuss the dependence of the Majorana hybridization energy $\varepsilon$ on
various parameters in realistic settings. By way of example, we consider a Bi$_2$Se$_3$
nanowire, with Fermi
velocities $v_2= 1.47v_1$ and $\hbar v_1=226$~meV$\times$nm \cite{Egger2010}. Choosing the nanowire radius $R=35$~nm in the outer regions and 
a constriction of radius $R_0=R/2$, a large size quantization gap $M_0 \simeq
7.14$~meV opens up in the central region.  Assuming a 
proximity gap $\Delta = 0.18$~meV in the outer regions, Fig.~\ref{figure2} shows $\varepsilon=\varepsilon(0)$ as  function of the constriction length $W$ for 
several values of the electrochemical potential $\mu$.  
The shown results, which have been obtained by numerical solution of Eq.~\eqref{deter}, 
are consistent  with the exponential scaling $\varepsilon\sim e^{-W/\xi}$ for $W\gg \xi$, see Eq.~\eqref{exposcal}. Moreover, the length scale $\xi=\xi(\mu)$ 
extracted from Fig.~\ref{figure2} also agrees with the prediction in 
Eq.~\eqref{xidef}. We observe from Fig.~\ref{figure2}  that
 for a short constriction,  the Majorana states move to high energies and eventually 
 approach the continuum part of the spectrum,  $\varepsilon\to \Delta$, for $W\to 0$.
 
With increasing local electrochemical potential $\mu$ of the central region, the transparency of the weak link, and hence the hybridization $\varepsilon$, will
also increase. 
This trend is visible in Fig.~\ref{figure2} and 
suggests that $\varepsilon$ may be changed in a 
convenient manner by gating the constriction and thereby tuning $\mu$.
(The minimal hybridization for given $W$ is reached for $\mu=0$.)
Fig.~\ref{figure3} shows in more detail how changes in $\mu$ will affect
the hybridization, see also Sec.~\ref{sec3d}. These numerical results nicely match the analytical predictions  
in Eqs.~\eqref{exposcal} and \eqref{xidef}.

\section{Majorana box qubits}\label{sec3}

The finite-length TI nanoribbon device in Fig.~\ref{figure1} 
supports four MBSs  near the ends of the proximitized regions. 
These states  correspond to Majorana fermion operators, 
$\gamma_j=\gamma_j^\dagger$, which 
 obey the Clifford anticommutator 
algebra $\{\gamma_j,\gamma_k\}=2\delta_{jk}$ \cite{Alicea2012}.  
Provided the proximitized parts are much longer than the length scale
$\hbar v_1/\Delta$, the outer MBSs ($\gamma_1$ and
$\gamma_4$) effectively represent zero modes, 
and the low-energy physics is governed by the Hamiltonian
 \begin{equation}\label{heff}
 H_{\rm eff}=i\varepsilon\gamma_2\gamma_3+H_C, 
\end{equation}
containing the hybridization energy $\varepsilon$ between $\gamma_2$ and $\gamma_3$.  The additional term $H_C$ describes charging and/or Josephson energies, which may be engineered on top of the basic setup discussed above. This generalization is introduced in Sec.~\ref{sec3a} before
we show in Sec.~\ref{sec3b} how it is key to the encoding of topologically protected 
qubits  in the Majorana Hilbert space.     
Inspired by recent Majorana box qubit proposals (tailor-made for the 
semiconductor-based  architecture) \cite{Plugge2017,Karzig2017}, we outline   
in Sec.~\ref{sec3c} the design of proof-of-principle experiments  testing the usefulness of
the TI nanoribbon platform for elementary QIP operations.  
(An alternative approach suggested in Ref.~\cite{Aasen2016} is to implement the basic 
anyon fusion protocols required for Majorana braiding operations, see Sec.~\ref{sec3a}.) 
We then discuss  the TI-based Majorana box qubit and compare it 
to other implementations in Sec.~\ref{sec3d}.
 Long-term perspectives of the current design include realizations of Majorana surface codes \cite{Landau2016,Plugge2016} via arrays of Majorana box qubits. We briefly discuss layouts of this type  in Sec.~\ref{sec4}.

\begin{figure}
\centering
\includegraphics[width=0.7\columnwidth]{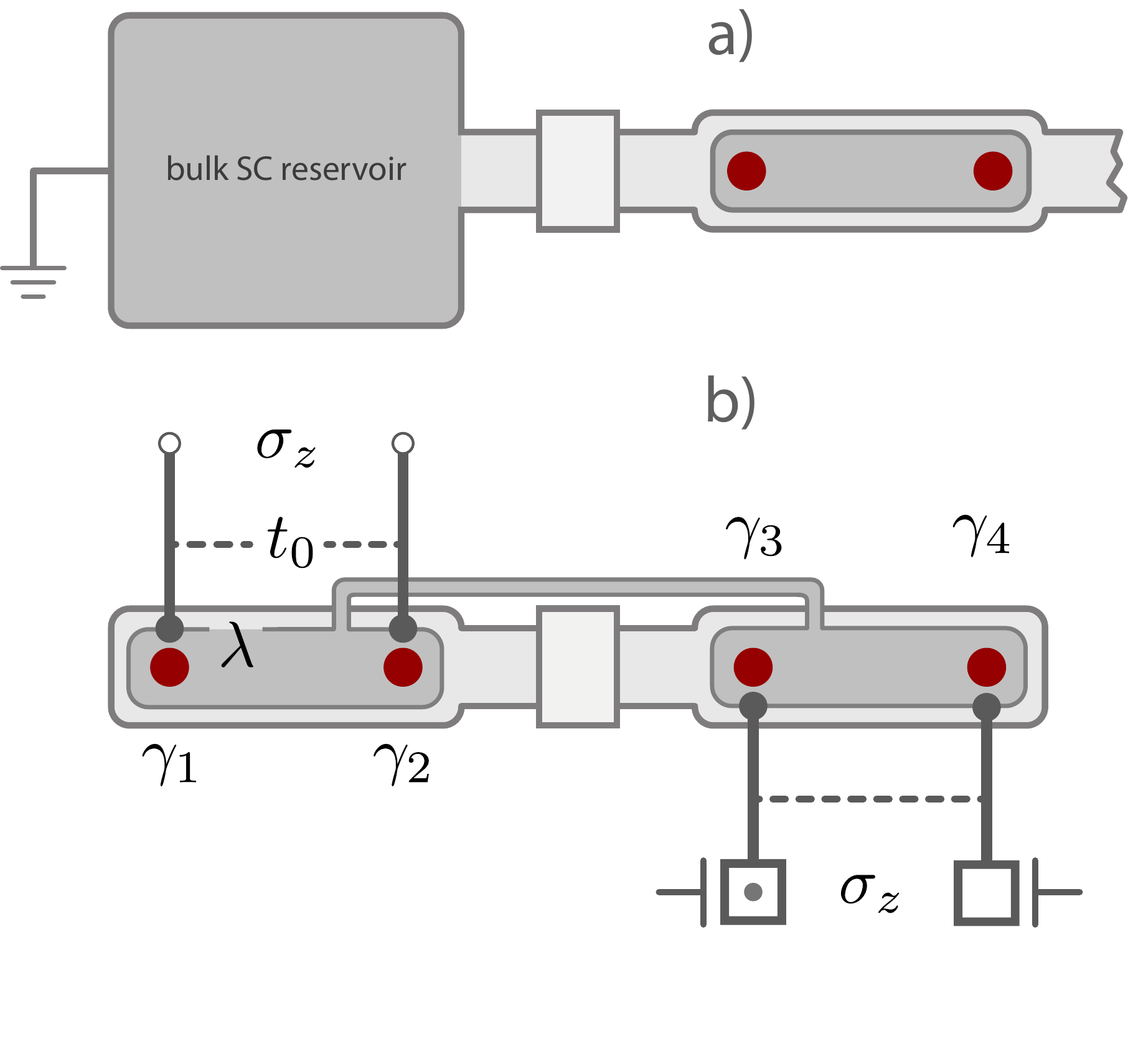}
\caption{Grounded vs floating devices. (a) Switchable grounding of TI nanoribbon devices: By tuning the gate voltage applied to a local top gate at the 
narrowed section, the Josephson coupling energy $E_J$ between the TI nanoribbon and a
grounded superconducting reservoir can be changed. As a consequence, one can switch
between a grounded and a floating device. (b) Floating version of the device in
Fig.~\ref{figure1}. The two $s$-wave superconductors are connected by a
superconducting bridge, i.e., the Majorana box is characterized by a single charging
energy. Normal leads (vertical black lines)  are tunnel-coupled to
individual MBSs and amongst themselves by interference links
to allow for interferometric readout of Pauli operators. The external boxes are symbolic for  
quantum dots or single-electron transistors which can be used  to pump single electrons
through the device and manipulate or read out  qubit states. For details, see main
text.  }
\label{figure4}
\end{figure}

\subsection{Floating vs grounded device}\label{sec3a}

Braiding protocols for MBSs require  switchable grounding of
the host device \cite{Aasen2016}, i.e., the option to isolate the system against
ground such that its finite capacitance defines an effective charging energy $E_C$.
The principle  is illustrated in Fig.~\ref{figure4}(a), which
differs from Fig.~\ref{figure1} in that the connector to ground is replaced by a
superconducting reservoir coupled to the system via a narrow TI wire segment.  As
with the  weak link in Sec.~\ref{sec2}, the geometric confinement implies that the
connecting section is gapped and effectively realizes a tunnel junction. A  top gate
may be installed to tune the tunneling strength and, thereby, the Josephson energy
$E_J$ between the superconducting regions connected by the junction. In this way,  changes in the gate voltage effect a switch between floating
(small $E_J$) and grounded  (large $E_J$) configurations,
see~Ref.~\cite{Aasen2016} for details.

The energy balance of a topological superconductor generally contains a charging
energy, $E_C$, and a dimensionless backgate parameter
$n_g$~\cite{Fu2010,Beri2012,Altland2013,Beri2013} controlling the energetically
preferred charge on it. Here, $E_C$ relates to the electrostatic capacitance of those
regions of the device that are in good electrical contact with each other. For
instance, the floating device shown in Fig.~\ref{figure4}(b) contains a
superconducting bridge connecting its left and right half, and this leads to a
`Majorana box' characterized by a single charging energy. In general, the  energy
balance is influenced by both, Josephson coupling, $E_J$, to a bulk superconductor
[as in Fig.~\ref{figure4}(a)], and a charging energy, $E_C$, and the sum of these
contributions,
\begin{equation}\label{HC}
H_C= E_C ( 2\hat N_s+ \hat n_\gamma -n_g)^2-E_J  \cos \hat \phi_s 
\end{equation}
adds to the Hamiltonian $H_{\rm eff}$ in Eq.~(\ref{heff})
\cite{Fu2010,Beri2012,Altland2013,Beri2013}. The number operator $\hat
N_s$ is canonically conjugate to the phase difference between the superconductors, $\hat \phi_s$, and counts the number of Cooper pairs on
the island, while $\hat n_\gamma= (i/2)[\gamma_1\gamma_2+\gamma_3\gamma_4]$ measures
the fermion number in the Majorana sector.  The Hamiltonian  $H_{\rm eff}$ describes
the low-energy sector of the system in that  above-gap quasiparticles and 
surface states with $j\ne -1/2$ are not taken into account. 

Externally imposed changes in the ratio $E_J/E_C$ may be applied to access Majorana
fermion occupancies, e.g., via the parity-to-charge conversion protocol of
Ref.~\cite{Aasen2016}. For few-$\mu$m long nanowires, a typical charging energy is
$E_C\approx 0.1$~K. We then expect a tunable parameter range of order $0.1\alt
E_J/E_C\alt 10$. This tunability is of key importance to  QIP protocols relying on
the non-Abelian braiding statistics of MBSs \cite{Aasen2016}.
 
\subsection{Majorana box qubit}\label{sec3b}

We next consider the floating device (`box') shown in Fig.~\ref{figure4}(b), where
$E_J=0$ and a Majorana box qubit can be realized. For a long TI nanoribbon with
exponentially small hybridization energy, $\varepsilon$,  the four MBSs in
Fig.~\ref{figure4}(b) effectively represent zero-energy modes. On energy scales small
against both $E_C$ and the proximity gap $\Delta$, and assuming that the backgate
parameter $n_g$ is close to an integer value, charge quantization on the box implies
that fermion parity is a good quantum number, $\gamma_1\gamma_2
\gamma_3\gamma_4={\cal P}=\pm 1$.  As a consequence,
the Majorana box has a two-fold degenerate ground state.  
The different components of the corresponding emergent spin-$1/2$
operator are encoded by spatially separated Majorana operators.  
Specifically, we choose Pauli operators as \cite{Beri2012,Altland2013,Beri2013} 
\begin{equation}\label{spindef}
 \sigma_x=i\gamma_3\gamma_1, \quad
 \quad \sigma_y=i\gamma_2 \gamma_3, \quad \sigma_z= i\gamma_1\gamma_2.
\end{equation}
Equivalent representations follow from the parity constraint, e.g., 
$\sigma_z=-i{\cal P}\gamma_3\gamma_4$, see Fig.~\ref{figure4}(b).

The degenerate two-level system defined in Eq.~\eqref{spindef} can effectively encode arbitrary qubit states 
$|\psi\rangle=\alpha|0\rangle+\beta |1\rangle$ (with
 $\sigma_z|0\rangle=|0\rangle$ and $\sigma_z|1\rangle=-|1\rangle$).  
The option to address different Pauli operators via spatially non-local access 
operations in a topologically protected setting, cf.~Sec.~\ref{sec3c},  holds promise for 
 a versatile and robust hardware qubit for QIP applications, see Refs.~\cite{Landau2016,Plugge2016,Plugge2017,Karzig2017}. 
 We expect that  at temperatures of a few milli-Kelvin,
  above-gap quasiparticles will limit the qubit lifetime. Although further work would be required to quantify the expected time scales, we note that the mechanisms for decoherence (quasiparticle poisoning, in the first place) are similar to those in
other topological Majorana qubits \cite{Aasen2016,Plugge2017,Karzig2017,Plugge2016}. 
   In what follows, we consider protocols operating on 
 shorter time scales where such detrimental effects can be neglected.

\subsection{Basic quantum operations} \label{sec3c}

By suitably designing the device, Pauli operators can be read out via interferometric
conductance measurements~\cite{Plugge2016,Plugge2017}.  
For the Majorana box qubit in Fig.~\ref{figure4}(b), these measurements would require the coupling of a pair of MBSs to tunnel electrodes.  
For instance, the coupling to $\gamma_1$ and $\gamma_2$ 
would amount to addressing
the  Pauli operator $\sigma_z$ of Eq.~\eqref{spindef}. Consider these two access leads
connected by an additional interference link of tunnel strength $t_0$ away from the
device (`reference arm'). Electron transport from lead $1\to 2$ can then either be (i)
through the box, where the cotunneling amplitude is given by 
$it_z\gamma_1\gamma_2=t_z \sigma_z$ with $t_z\simeq \lambda^2/E_C$ 
for elementary tunnel amplitude $\lambda$, or (ii) 
through the reference arm with amplitude $t_0$.
The tunnel conductance between leads 1 and 2 is thus given by
\begin{equation}
G_{12} = \frac{e^2}{h} \nu_1\nu_2 |t_0+ t_z \sigma|^2,
\end{equation}
where $\sigma=\pm$ refers to the  eigenvalues of $\sigma_z$ and $\nu_{1,2}$ denotes
the respective density of states in the normal leads. The outcome of the
measurement  depends on $\sigma$, and this means that  the conductance measurement
projects the original qubit state $|\psi\rangle=\alpha|0\rangle+\beta|1\rangle$ to
the respective $\sigma_z$-eigenstate. With probability $|\alpha|^2$
($|\beta|^2=1-|\alpha|^2$) the qubit assumes the state $|0\rangle$ ($|1\rangle$)
after the measurement. Projective conductance measurements of this type may be
applied to read out arbitrary Pauli operators or to  initialize the
qubit in a Pauli eigenstate.  It is worth mentioning that for the device in Fig.~\ref{figure4}(b), the Pauli operator $\sigma_y=i\gamma_2\gamma_3$ can 
be measured without an additional reference link shared by leads 2 and 3 
since the central superconducting bridge already provides this
link \cite{Karzig2017}.

The controlled \emph{manipulation} of qubit states, $|\psi\rangle$, however, requires additional access elements. Specifically, the application of a given Pauli operator~\eqref{spindef}
to a state $|\psi\rangle$ can be realized through the pumping of a single electron between a pair of quantum dots (or single-electron transistors) tunnel-coupled to the  MBSs corresponding to the operator~\cite{Landau2016,Plugge2016,Plugge2017}.  Here, the dots are assumed to be in the single-occupancy regime, 
where the respective energy levels can be changed by means of gate voltages. 
The pumping of a single electron from dot $1\to 2$ then  implies 
a unitary state transformation, $|\psi\rangle\to U|\psi\rangle$, where $U$ corresponds 
to the respective Pauli operator.  This transformation law is topologically robust in
the sense that it is independent of details of the protocol 
 \cite{Plugge2017}. For instance, 
it does not depend on the values of the tunnel couplings
nor on the precise time dependence of gate voltages. 
However, it has to be made sure that the electron ends up in the
desired final state,
either by a confirmation measurement of the dot charge or by 
running the protocol in an adiabatically slow fashion.
Finally, also pairs of dots may be 
connected by additional phase-coherent reference arms to effectively realize 
arbitrary single-qubit phase gates.
However, in contrast to the Pauli operators discussed above, such 
phase gates generally are not topologically protected anymore 
 due to measurement-induced dephasing processes,
 see~Ref.~\cite{Plugge2017} for details. The latter processes are
 essential for projective readout and/or initialization 
 but are detrimental to manipulations of the qubit state.

Alternative proposals to read out, manipulate, or initialize Majorana box qubit
states can be found in Refs.~\cite{Plugge2017,Karzig2017}.  For example, quantum dots
may be applied as an alternative to leads for readout purposes.
Furthermore, by measuring products of Pauli operators on different boxes,
one may entangle the states of the corresponding qubits. In particular,
a measurement-based protocol for generating a controlled-NOT gate can be found
in Ref.~\cite{Plugge2017}.

\subsection{Discussion}\label{sec3d}

We now turn to a critical discussion of the proposed platform.
Since MBSs in the TI setup are formed from topological surface states, they can be expected to
enjoy an intrinsic protection mechanism against both elastic impurity scattering  and pair-breaking disorder.  
At the same time, presently available TI materials are not as clean as the corresponding semiconductor nanowire systems,
and significant experimental progress will be needed to verify the practical usefulness of this platform.
 In what follows, we address the robustness of MBSs in TI nanoribbon devices against possibly detrimental mechanisms
 in view of the above results, and compare the TI implementation to alternative realizations.    

First, it may be difficult to precisely tune the magnetic flux 
to $\varphi=1/2$ in the proximitized regions, even though one can adjust 
magnetic fields to high accuracy.  Such a flux mismatch could arise 
because of  (i) inhomogeneities in the cross section area, (ii) misalignment 
between the magnetic field and the TI nanoribbon axis which, in addition, 
weakly breaks rotation symmetry and therefore mixes $j=-1/2$ states 
with $j\ne -1/2$ high-energy states, and/or (iii) because different 
nanowire parts may not be exactly parallel to each other.   For  small flux mismatch,
the topological energy gap appearing in Eq.~\eqref{spectr0} will change only slightly  
due to the corresponding change in $M(\varphi)$, see Eq.~\eqref{massgap},
without affecting the robustness of MBSs.
However, one then needs a finite  electrochemical potential 
$\mu_S=\mu(|z|>W/2)$ in the proximitized regions, in contrast to 
our assumption in Eq.~\eqref{delmu}.
In particular, $M(\varphi)< |\mu_S|<\hbar v_2/R$ 
is required for well-defined helical 1D states when
$\Delta=0$ (which then yield MBSs for $\Delta\ne 0$).
We conclude that flux mismatch is not expected to create serious
problems for the robustness of MBSs.
 
Second, we address what happens for finite electrochemical potential $\mu_S$ in the proximitized regions, where we focus on the case $\varphi=1/2$. 
 Recalling that the $\Delta=0$ states with 
 $j=-1/2$ have linear dispersion, we expect that  
 a shift of $\mu_S$ enters physical quantities mainly through
 the difference $\mu-\mu_S$. In effect, the above $\mu_S=0$ results thus 
  apply again.

Next we consider the localization length of MBSs in our setup and compare the
result to other platforms.  For the device in Fig.~\ref{figure1},  different length
scales govern the MBS decay into the inner  and the outer part.  Taking the proximity gap as $\Delta=0.18$~meV, 
one gets  $\xi_\Delta=\hbar v_1/\Delta=1.25~\mu$m on the superconducting side,
while the decay into the inner segment is governed by the much
shorter length $\xi$ in Eq.~\eqref{xidef}.  (The precise value of $\xi$ depends
on the parameters $W,R_0/R,$ and $\mu$.) The MBS localization length is longer than $\xi_\Delta$ in 
semiconductor nanowires, where a typical spin-orbit coupling energy $\hbar \alpha\approx 20$~eV$\times$nm translates into the length 
scale $125~$nm \cite{Sarma2012}. 
The above estimates indicate that one may need rather long proximitized 
TI nanowires (exceeding at least $5~\mu$m) in order to have negligible MBS 
overlap, e.g., between  $\gamma_1$ and $\gamma_2$ in Fig.~\ref{figure4}(b). 
While this requirement constitutes a slight disadvantage against 
semiconductor nanowires, we note that sufficiently long TI nanoribbons are 
already available \cite{Cho2015,Jauregui2016}.  Similar values for the MBS localization length as found for
the TI case above have also been estimated for the HgTe platform \cite{Deacon2017,Bocquillon2017}.

We continue by studying the maximum time scale $t_0$ on which the simplest 
type of Majorana qubit could be operated without dephasing for different platforms, using a device 
as in Fig.~\ref{figure1}. In our TI setting, this scale is defined by 
$t^{-1}_0=\varepsilon(0)$, see Eq.~\eqref{exposcal}.
Indeed, for times $t>t_0$, the hybridization of the inner MBSs in 
Fig.~\ref{figure1} will inevitably dephase the qubit state. 
Clearly, it is then desirable to access time scales $t_0$ as long as possible. 
We observe from Eq.~\eqref{xidef} that this condition is reached by choosing
$\mu=0$ and a large width $W$ of the central segment, see Figs.~\ref{figure2}
and \ref{figure3}.  For instance, choosing $W=300$~nm and $|\mu|<0.2$~meV,
we find $\varepsilon(0)<0.027~\mu$eV, and hence $t_0>2.4~\mu$s.  
This time scale exceeds the one estimated for semiconductor 
nanowires \cite{Aasen2016}.  However, in practice an important additional limitation on operation
times for Majorana qubits may come from quasiparticle poisoning.  The poisoning time is known to
be $\agt 1~\mu$s for semiconductor Majorana devices \cite{Albrecht2016b} but 
remains to be studied for proximitized TI systems.

\begin{figure}[t]
\centering
\includegraphics[width=0.7\columnwidth]{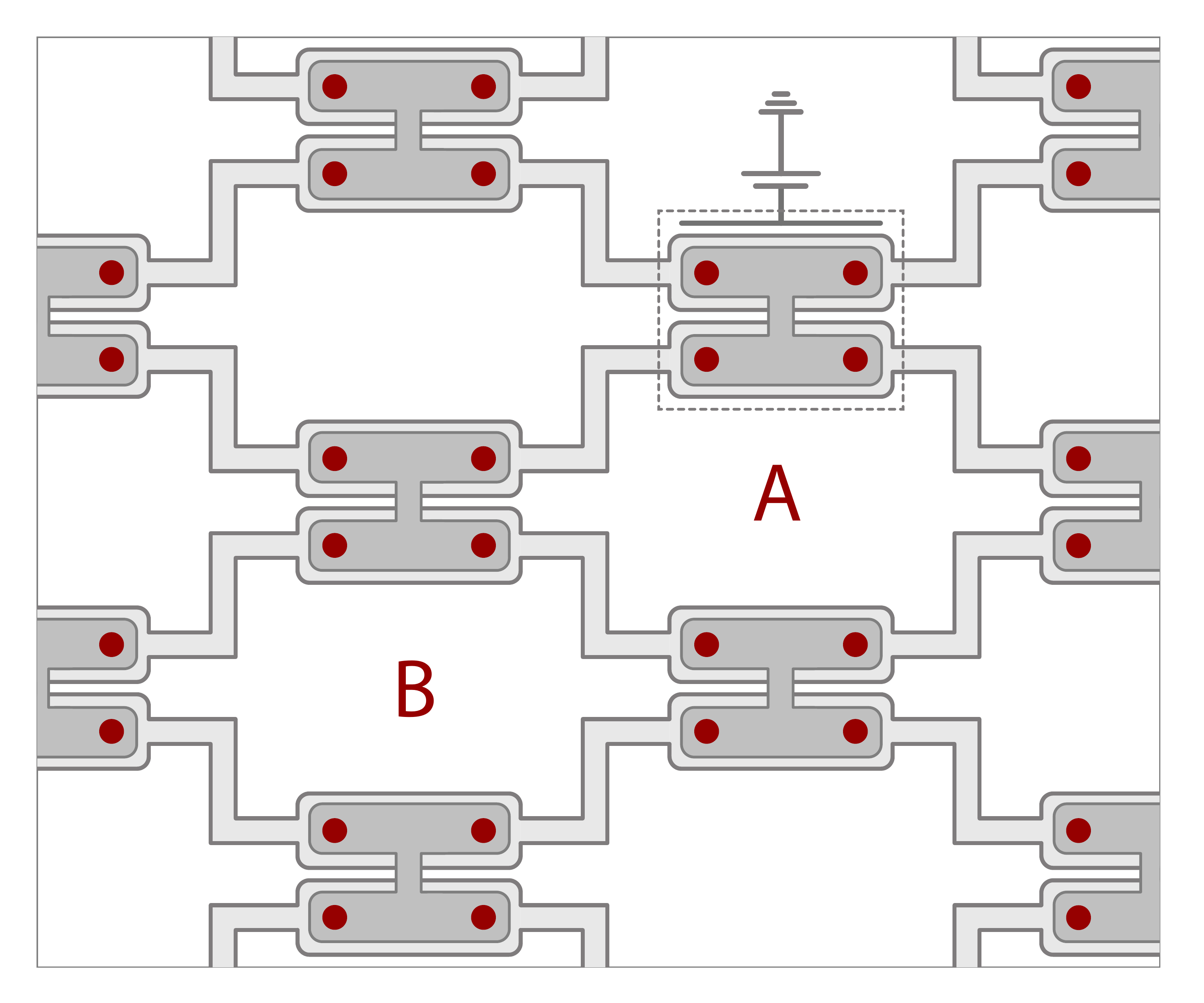}
\caption{ 2D network of Majorana box qubits using proximitized TI nanoribbons for
implementing a Majorana surface code.  Stabilizers of type A or B correspond to 
products of eight Majorana operators around a minimal plaquette as indicated.
Access elements for initialization, manipulation, and readout of stabilizers 
are not shown but described in the main text.  }\label{figure5}
\end{figure}

As elaborated in Sec.~\ref{sec1}, semiconductor nanowires 
 define the experimentally most advanced platform for Majorana states at present, 
 and detailed proposals for Majorana qubits in that platform 
 have appeared \cite{Vijay2016,Aasen2016,Plugge2017,Karzig2017}. 
Nonetheless, this implementation also has some drawbacks.
As remarked before, the chemical potential has to be chosen close to the 
band bottom, which in turn renders states susceptible to the effects of disorder. 
 Such problems could probably be avoided in a 2D architecture, where a 
 2D electron gas (2DEG) with strong spin-orbit coupling is proximitized by a 
 lithographically patterned superconducting top layer \cite{Shabani2016,Hell2017,Pientka2017}.  
 Nonetheless, the chemical potential window allowing for robust MBSs is arguably bigger
  for the TI nanoribbon case.
 Moreover, the implementation of Majorana box qubits in a semiconductor
 nanowire setting has encountered difficulties due to the need for separate
 reference links \cite{Marcus2017}.   Using our TI nanoribbon setup (and similarly for 2DEG implementations),
 this problem can be avoided by designing reference links from the TI itself,  see Figs.~\ref{figure4} and \ref{figure5}.

\section{Conclusions and outlook}\label{sec4}

In this paper, we have outlined how Majorana qubits can be defined and operated in a
proximitized TI nanoribbon architecture.  The key element of the construction are
gate-tunable internal tunnel junctions realized through narrowed regions of lowered  axial magnetic flux
in TI nanoribbons. 
This allows us to tune the hybridization of MBSs emerging from a topologically protected helical 1D surface state mode, and thereby makes it possible to manipulate the quantum information. 
The linear dispersion of the 1D modes in this platform is expected to give the MBSs a high level of robustness. (In this regard, the situation may be better than  in semiconductor wire platforms, where one is operating close to the bottom of a parabolic band.) We are
confident that this platform is sufficiently versatile and flexible to implement the 
 quantum information processing protocols outlined above.   

 Once proof-of-principle experiments
have confirmed this expectation, one may envisage the extension of the system to 2D networks containing
many Majorana box qubits.  
Specifically, the blueprint sketched in Fig.~\ref{figure5} indicates the extension to a network realizing a two-dimensional surface code, 
cf.~Refs.~\cite{Terhal2012,Vijay2015,Landau2016,Plugge2016}.
The surface code approach builds on so-called stabilizer operators  
corresponding to products of eight Majorana operators surrounding the minimal plaquettes of Fig.~\ref{figure5}.  There are two types (A and B) of such operators, 
and the essence of the surface code is that all of them
commute. The binary eigenvalues $\pm 1$ of the stabilizers then define the physical qubits of the system. During each operational cycle of the system the majority of these qubits is measured (`stabilized'), and projection onto the highly entangled degenerate ground states of the system takes place. The few qubits exempt from the measurement process serve as logical qubits and can be manipulated along the lines of the discussion above.

In contrast to the `unfolded' linearly arranged box qubit in Fig.~\ref{figure4}(b), Fig.~\ref{figure5} suggests an alternative construction, where pairs of adjacent proximitized TI nanowires are connected
through superconducting bridges to form 90-deg rotated `H'-type structures.
Each of these structures represents a Majorana box with its own charging
energy. Pairs of adjacent MBSs on neighboring boxes are connected by tunnel 
links as shown in Fig.~\ref{figure5}.  Since the TI nanoribbon network is most likely
fabricated by lithographic and/or wet etching means, the present platform 
would naturally employ tunable TI nanoribbon parts for those tunnel links as well.
In this way,  the need for separate wires and/or 
other materials required by the corresponding semiconductor architecture \cite{Plugge2016,Plugge2017,Karzig2017} might be avoided.  Since the proximitized TI nanoribbon parts in  Fig.~\ref{figure5} are arranged parallel to each other, the MBSs can be generated simultaneously under a uniform applied magnetic field,
provided the nanoribbon cross section can be accurately controlled in the 
fabrication process.  

\acknowledgments 
We wish to thank Stephan Plugge for discussions.
Funding by the Deutsche Forschungsgemeinschaft (Bonn) 
within the networks CRC TR 183 (project C04) and CRC 1238 (project A04)
is acknowledged.

\appendix
\section{Spinor wave functions}\label{appa}

Here we provide the explicit form of the BdG  
eigenstates, $\Psi(z)$,  for the device in Fig.~\ref{figure1}.  For notational 
simplicity, we employ units with $\hbar v_1=1$.
Putting $j=-1/2$, since we are interested in constructing 
Majorana bound states, we consider 
only energies below the superconducting gap, $|E|<\Delta$. 
With $H_{\rm BdG}$ in Eq.~\eqref{Ham0},
we shall first write down general solutions of the BdG equation 
in each of the three regions. These solutions are subsequently
matched at the interface points $z=\pm W/2$ by continuity.
Using parameters $A_{1,2}^{(\pm)}$, the solution for $|z|>W/2$ 
decaying at $|z|\to \infty$ reads
\begin{widetext}
\begin{equation}\label{eqA1} 
\Psi(z)\Bigr|_{|z|>W/2} =  e^{-\sqrt{\Delta^2-E^2}|z|} \left[ A^{(s)}_1 
\left(\begin{array}{c} E \\ -s\sqrt{\Delta^2-E^2}  \\ 0 \\ \Delta e^{is\phi/2} 
\end{array}\right) 
 + A^{(s)}_2 \left(\begin{array}{c} -s\sqrt{\Delta^2-E^2}  
 \\-E\\ \Delta e^{is\phi/2} \\ 0 \end{array}\right) \right],\quad s={\rm sgn}(z)=\pm,
\end{equation}
where the first and second (third and fourth) component refers to the spin structure 
of the particle (hole) part of the Nambu spinor. 
In the central region $|z|<W/2$, with coefficients $B^{(\pm)}_{1,2}$ the 
solution is given by 
\begin{equation}\label{eqA2}
 \Psi(z) \Bigr|_{|z|<W/2}=   \sum_{\pm}
 \left[ B^{(\pm)}_1  e^{\pm\sqrt{M_0^2-(E+\mu)^2}z} 
\left(\begin{array}{c} \pm(M_0+E+\mu) \\ \sqrt{M_0^2-(E+\mu)^2}
 \\ 0 \\ 0 \end{array}\right) 
 +  B^{(\pm)}_2  e^{\pm\sqrt{M_0^2-(E-\mu)^2}z} 
\left(\begin{array}{c} 0\\ 0\\ \pm(M_0-E+\mu) \\ \sqrt{M_0^2-(E-\mu)^2}
 \end{array}\right) \right].
\end{equation}
Imposing continuity at $z=\pm W/2$, we find that eigenenergies 
with $|E|< {\rm min}(\Delta,M_0)$ follow from the zero-determinant condition in Eq.~\eqref{deter}. The determinant $D(E)$ is a symmetric function of $E$ and given by
\begin{eqnarray}\label{Ddef}
D(E) &=& -2\Delta^2 a_+ a_- \cos\phi + \sum_{\pm}  \left[ (2E^2-\Delta^2)
\left (a_+ a_- \pm \mu^2\mp E^2\right)
\pm M_0^2\Delta^2\right] \cosh[(a_-\pm a_+)W]  \\ &+&
 2E\sqrt{\Delta^2-E^2} \sum_\pm \left[ (E - \mu) a_+ \pm (E+ \mu) a_- \right] \sinh
 [(a_-\pm a_+)W] , \nonumber \quad a_\pm (E) \equiv \sqrt{M_0^2-(E\pm \mu)^2}.
\end{eqnarray}
\end{widetext}

\end{document}